\documentclass[aps,pre,twocolumn,amsmath,amssymb,superscriptaddress]{revtex4-2}

\usepackage{mathrsfs}
\usepackage{graphicx}
\usepackage{bm}
\usepackage{threeparttable}
\usepackage{amsmath,amssymb}
\usepackage{color}
\usepackage{tikz}
\usepackage{epstopdf}
\usepackage{multirow}
\usepackage[colorlinks=true,linkcolor=blue,anchorcolor=blue, citecolor=blue,urlcolor=blue]{hyperref}

\def\h{h}
\def\tmd{\Delta t_{MD}}
\def\a{\alpha}
\def\b{\beta}
\def\g{\gamma}
\def\pd#1{\partial_{#1}}
\def\kr#1{\delta_{#1}}
\def\ep#1{\varepsilon_{#1}}
\def\n{\Lambda}

\begin{document}

\preprint{}

\title{{Colloidal phoresis in two-dimensional odd fluids}}

\author{Yuxing Jiao}
\email{jiaoyuxing22@mails.ucas.ac.cn}
\affiliation{Beijing National Laboratory for Condensed Matter Physics and Laboratory of Soft Matter Physics, Institute of Physics, Chinese Academy of Sciences, Beijing 100190, China}\affiliation{School of Physical Sciences, University of Chinese Academy of Sciences, Beijing 100049, China}

\author{Qing Yang}
\affiliation{Department of Physics, Wenzhou University, Wenzhou, Zhejiang 325035, China}
\affiliation{Beijing National Laboratory for Condensed Matter Physics and Laboratory of Soft Matter Physics, Institute of Physics, Chinese Academy of Sciences, Beijing 100190, China}

\author{Mingcheng Yang}
\email{mcyang@iphy.ac.cn}
\affiliation{Beijing National Laboratory for Condensed Matter Physics and Laboratory of Soft Matter Physics, Institute of Physics, Chinese Academy of Sciences, Beijing 100190, China}\affiliation{School of Physical Sciences, University of Chinese Academy of Sciences, Beijing 100049, China}

\begin{abstract}
    Under a thermodynamic gradient, for example, the concentration or temperature gradients, the colloidal particles immersed in the solvent can exhibit a directional migration along or against the gradient---phoresis, a cross transport effect. When the solvent is an odd fluid, where the time-reversal and parity symmetries are broken microscopically, the odd transport phenomenon is allowed. This means an odd phoresis may appear: the colloidal particle migrates perpendicularly to the thermodynamic gradient. Here, we realize the odd diffusiophoresis and odd thermophoresis for a colloidal particle immersed in a two-dimensional odd fluid by performing mesoscale fluid simulations. We quantify the odd phoretic factors and further provide the phoresis-osmotic flow field driven by a phoretic colloidal particle, which is quantitatively consistent with the numerical solutions of the corresponding odd fluid dynamics equations.
\end{abstract}

\maketitle

\section{Introduction}
Odd fluids are a class of fluids with non-vanishing odd transport coefficients. These coefficients are the anti-symmetric components in the transport coefficient tensor, and their existence is allowed by broken time-reversal symmetry of the odd fluids according to the Onsager-Casimir reciprocal relations~\cite{NET}. Typical examples for odd fluids include electron Hall fluids~\cite{Avron1995,Hoyos2012,Bandurin2019,Holder2019}, polyatomic gases in a magnetic field~\cite{PolyGas,PolyHD,PolyGas3_Dufor,PolyGas4_thermaldiffusion,PolyGas_thermaldiffusion2}, and chiral active fluids~\cite{ActiveColloidal,Vitelli_Fluhydro,Markovich2021,Hargus2021}. One striking feature of odd fluids brought by the odd transport coefficients is that thermodynamic gradients in an odd fluid can drive fluxes perpendicular to the gradients. As a result, odd fluids exhibit more intricate and intriguing transport behaviors compared to common fluids, making them a subject of widespread interest in recent years. For instance, in the presence of concentration, momentum, and temperature gradients, odd fluids, respectively, display transverse conjugate mass, momentum and heat fluxes~\cite{Hargus2021,Banerjee2017,PolyGas,PolyHD}. Additionally, odd fluids exhibit a more counterintuitive phenomenon known as the odd cross transport effect. In this effect, fluxes, including transverse fluxes, can be generated by non-conjugate gradients. The prominent examples include the transverse thermal diffusion effect~\cite{PolyGas4_thermaldiffusion,PolyGas_thermaldiffusion2} (Soret effect) and the transverse diffusion-thermo effect~\cite{PolyGas3_Dufor} (Dufour effect), which have been observed in molecular mixtures of polyatomic gases (e.g., the $\text{N}_2$-Ar mixture~\cite{PolyGas3_Dufor,PolyGas4_thermaldiffusion}) under a magnetic field.

Besides odd molecular mixtures, colloidal particles immersed in odd solutions could be expected to possess the odd phoresis effect, but that has never been reported. In general, phoresis refers to the migration of colloidal particles (or polymers) induced by thermodynamic gradients in a normal solvent~\cite{Anderson1989_Pho}. Depending on the type of gradient, this phenomenon has specific terms, such as diffusiophoresis~\cite{Anderson1984_DiffPho,MD_DiffPho_Maldarelli2013,Annunziata2024,Lechnik1984} (driven by concentration gradients) or thermophoresis~\cite{Piazza_2008_ThPho,Wurger_2010_ThPho,Parola2019,Yang2012,Yang2013,Yang2013_2} (driven by temperature gradients). From the perspective of non-equilibrium thermodynamics, diffusiophoresis can be treated as a cross-diffusion effect in a solute-solvent-colloid ternary system~\cite{Annunziata2024,Lechnik1984}, while thermophoresis corresponds to a thermal diffusion effect~\cite{Piazza_2008_ThPho}. Microscopically, the phoretic force that drives colloidal particles arises from the gradient-induced nonuniform interactions between the colloidal particles and the surrounding fluid particles. Remarkably, if the solvent is an odd fluid, thermodynamic gradients should additionally induce a transverse migration, which refers to the odd phoresis.

Hitherto, the normal phoresis transports have been intensively studied and widely used in the manipulation techniques of the mesoscale objects, such as colloidal separation and focusing in microfluidic devices~\cite{Abecassis_2009NJP,Piazza2010SM,Shin2020}, and the motion control of cells and biological molecules~\cite{Braun2006,Braun2010,Fukuyama2015,Fukuyama2020}. Besides phoresis itself, the flow field induced by the colloidal phoresis is also a notable topic in the microfluidic applications~\cite{Yang2013,Tan2019} and colloidal self-assembled techniques~\cite{Braun2008,Tomaso2017}. Compared to the normal phoresis, the odd phoresis, as a more intricate transport phenomenon, may therefore have considerable application prospects. However, studies on the odd phoresis, either in experiment or theory, are still lacking. Therefore, in this paper, we investigate the possible odd diffusiophoresis and odd thermophoresis of the colloidal particles in a two-dimensional (2D) odd fluid, by means of computer simulations. We find the colloidal particle experiences a significant odd phoretic force, and meanwhile its reaction force drives a transverse flow field around the colloidal particle. 

\section{The phoretic force}
A colloidal particle immersed in a fluid with a thermodynamic gradient can be subjected to a phoretic force, which results in the migration of the colloidal particle. {This phoretic force is an internal force that comes from the nonuniform distribution of solvent particles around the colloidal particle.}

In the linear response regime, this phoretic force can be generally expressed as a linear combination of the gradient. For example, the thermophoretic force $\bm{F}^{\rm{T}}$ is
\begin{equation}\label{FT}
    \bm{F}^{\rm{T}}=-k_B\bm{\alpha}^{\rm{T}}\cdot\nabla T,
\end{equation}
where $k_B$ is the Boltzmann constant and the dimensionless tensor $\bm{\alpha}^{\rm{T}}$ is the thermophoretic factor tensor. For a normal isotropic fluid, the classic Onsager reciprocal relation dictates this transport coefficient tensor is diagonal. However, 
for the odd fluid background, $\bm{\alpha}^{\rm{T}}$ takes the form of
\begin{equation}
    \bm{\alpha}^{\rm{T}}=\begin{bmatrix}
        \alpha^{\rm{T}}_e & \alpha^{\rm{T}}_o \\
        -\alpha^{\rm{T}}_o & \alpha^{\rm{T}}_e
    \end{bmatrix},
\end{equation}
where $\alpha^{\rm{T}}_e$ and $\alpha^{\rm{T}}_o$ are the regular and odd thermophoretic factor, respectively. Notice that $\alpha^{\rm{T}}_o$ is absent in the normal fluids. Similarly, the diffusiophoretic force $\bm{F}^{\rm{D}}$ is
\begin{equation}\label{FD}
    \bm{F}^{\rm{D}}=-\bm{\alpha}^{\rm{D}}\cdot\nabla\mu.
\end{equation}
Here, we consider a solute-solvent-colloid ternary system and denote the chemical potential of the solute by $\mu$. The diffusiophoretic factor tensor $\bm{\alpha}^{\rm{D}}$ is
\begin{equation}
    \bm{\alpha}^{\rm{D}}=\begin{bmatrix}
        \alpha^{\rm{D}}_e & \alpha^{\rm{D}}_o \\
        -\alpha^{\rm{D}}_o & \alpha^{\rm{D}}_e
    \end{bmatrix}.
\end{equation}

{We remark that there exist two different routes for measuring the phoretic factors~\cite{Lusebrink_2012}. The first is to study the migration velocity $\bm{U}$ of a freely moving colloidal particle under a certain thermodynamic gradient denoted by $\bm{G}$. The linear response relation between $\bm{U}$ and $\bm{G}$ reads
\begin{equation}
    {U}_{\alpha}={D}^p_{\alpha\beta}{G}_{\beta},
\end{equation}
where ${D}^p_{\alpha\beta}$ is the corresponding linear response coefficient, and $\bm{G}=-k_B\bm{\nabla}T$ for thermophoresis and $\bm{G}=-\bm{\nabla}\mu$ for diffusiophoresis. Multiplying both sides of this equation by the friction tensor $\bm{\gamma}$, we obtain the force balance condition:
\begin{equation}
    {\gamma}_{\alpha\beta}{U}_\beta={\gamma}_{\alpha\beta}{D}^p_{\beta\nu}{G}_\nu,
\end{equation}
where the left-hand side corresponds to the friction force ${f}_\alpha=-{\gamma}_{\alpha\beta}{U}_\beta$, and the right-hand side represents the phoretic force ${F}^p_\alpha={\gamma}_{\alpha\beta}{D}^p_{\beta\nu}{G}_\nu={\alpha}^p_{\alpha\beta}{G}_{\beta}$. Therefore, by measuring $\bm{U}$, one can calculate the phoretic factor via ${\alpha}^p_{\alpha\beta}={\gamma}_{\alpha\nu}{D}^p_{\nu\beta}$.

The second approach is to study a phoretic colloid constrained by an external force $\bm{F}$ in a thermodynamic gradient, for example, via a steep harmonic potential well~\cite{Galliero2008}. The force balance condition reads
\begin{equation}
    {F}_\alpha+{F}^p_\alpha={0}.
\end{equation}
Therefore, the phoretic force and the phoretic factor tensor can be directly determined by quantifying the external force $\bm{F}$.

Compared to the first approach, the second one is much more suitable for simulation studies with a finite simulation box size, where the migration range of the colloidal particle experiencing the same local environment is quite limited. We point out that this fixed-particle approach has been applied in simulation studies on normal thermophoretic colloidal particles to determine the thermophoretic force and thermophoretic factors~\cite{Yang2013}. Consequently, we utilize this second route in our study. More specifically, we study 2D odd diffusiophoresis and odd thermophoresis by simulating a colloidal particle fixed in a concentration gradient field and a temperature gradient field, respectively. 

In the following sections, we investigate the 2D odd phoresis from two aspects: (1) the transverse phoretic force, characterized by the odd phoretic factors, and (2) the accompanied 2D parity-violating flow pattern around the fixed phoretic particle.
}

\section{Simulation methods}\label{Methods}
Our simulation is performed by a hybrid mesoscopic-molecular dynamics scheme at the particle level. The motion of odd fluid particles is simulated by the chiral stochastic rotation dynamics (CSRD), an efficient mesoscale simulation model for odd fluids~\cite{CSRD}. This model was recently developed from stochastic rotation dynamics (SRD)~\cite{MPC1,MPC2,Kapral2008,Gompper_2009}, a well-established coarse-grained fluid simulation method. The interactions between the fluid particle and the colloidal particle are simulated by the standard molecular dynamics (MD)~\cite{MPC_MD}. 

\subsection{CSRD and hybrid MD scheme}
In CSRD model, the fluid consists of $N$ point particles with mass $m$ moving in  two-dimensional (2D) space. The positions $\bm{r}_i$ and velocities $\bm{v}_i$ ($i=1,\dots,N$) of these particles follow a time-discretized dynamics, which includes two main steps: streaming step and collision step. Firstly, in the streaming step, the particles move as free particles:
\begin{equation}\label{strEq}
    \bm{r}_i\left( t+\h \right)=\bm{r}_i(t)+\bm{v}_i(t)\h,
\end{equation}
where $\h$ is the discretized time step. Secondly, in the collision step, space is divided into a square lattice. To maintain the Galilean invariance, the position of the lattice is randomly shifted for every collision step~\cite{MPC_RS}. In each cell, fluid particles collide according to the rule:
\begin{equation}\label{colEq}
    \bm{v}_i\left( t+\h \right)=\bm{v}_{cm}+\bm{R}\cdot\left[\bm{v}_i(t)-\bm{v}_{cm}\right],
\end{equation}
where $\bm{v}_{cm}$ is the center-of-mass velocity of the cell and $\bm{R}=\bm{R}\left( \Omega+\theta \right)$ is a rotation matrix. Here, $\Omega$ is a stochastic rotation angle chosen from $\left\{ -\omega,\omega \right\}$ with equal probability and $\theta$ is a fixed chiral rotation angle. A nonzero $\theta$ breaks the time-reversal and parity symmetries of the dynamics, rendering nonvanishing odd transport coefficients including the odd self-diffusivity, odd viscosities, and odd heat conductivity available in the CSRD, and enabling the CSRD to correctly capture the hydrodynamics of odd fluids~\cite{CSRD}.

When a disk-shaped colloidal particle is fixed in the CSRD fluid, we apply a hybrid MD scheme~\cite{MPC_MD}. The interaction between the colloidal particle and fluid particles is modeled using the following repulsive Weeks-Chandler-Andersen (WCA) potential $\varphi$:
{\begin{equation}
    \varphi_n\left( r \right)=\begin{cases}
        4\epsilon\left[ \left( \frac{\sigma}{r} \right)^{2n}-\left( \frac{\sigma}{r} \right)^{n} +\frac{1}{4} \right], &     r\leqslant 2^{1/n}\sigma, \\
        0, & r > 2^{1/n}\sigma.
      \end{cases}
\end{equation}}where $\epsilon$ is the colloid-fluid energy scale and $\sigma$ is the colloid-fluid collision diameter. The specific values of these interaction potential parameters will be provided later. Now, in the streaming step, the motion equations of the fluid particles around the fixed colloidal particle are solved by using the velocity Verlet algorithm with a time step $\tmd$.

In our simulations, the units of mass, length, and energy are defined by the mass of fluid particle $m=1$, size of a cell $l=1$, and $k_B\overline{T}=1$, where $k_B=1$ and $\overline{T}$ is the average temperature of the simulation system. The parameters for the CSRD-MD hybrid scheme are set as follows: $\epsilon=2.5$, $\sigma=4$, $\h=50\tmd=0.1$, $\theta=5\pi/9$, $\omega=2\pi/3$ for diffusiophoresis, and $\omega=5\pi/6$ for thermophoresis.

\subsection{The setup of diffusiophoresis}
In the simulation of diffusiophoresis, we introduce two kinds of fluid particles, $A$ and $B$. We regard $A$ and $B$ as the molecular solute and the solvent, respectively. These particles obey the same collision rule, i.e., Eq.~\eqref{colEq}, and have the same mass $m$, but they interact differently with the colloidal particle. The interactions of $A$ particles and $B$ particles are, respectively, $\varphi_A\left( r \right)=\varphi_3\left( r \right)$ and $\varphi_B\left( r \right)=\varphi_{24}\left( r \right)$.

\begin{figure}[h]
    \centering
    \includegraphics[keepaspectratio, width=\columnwidth]{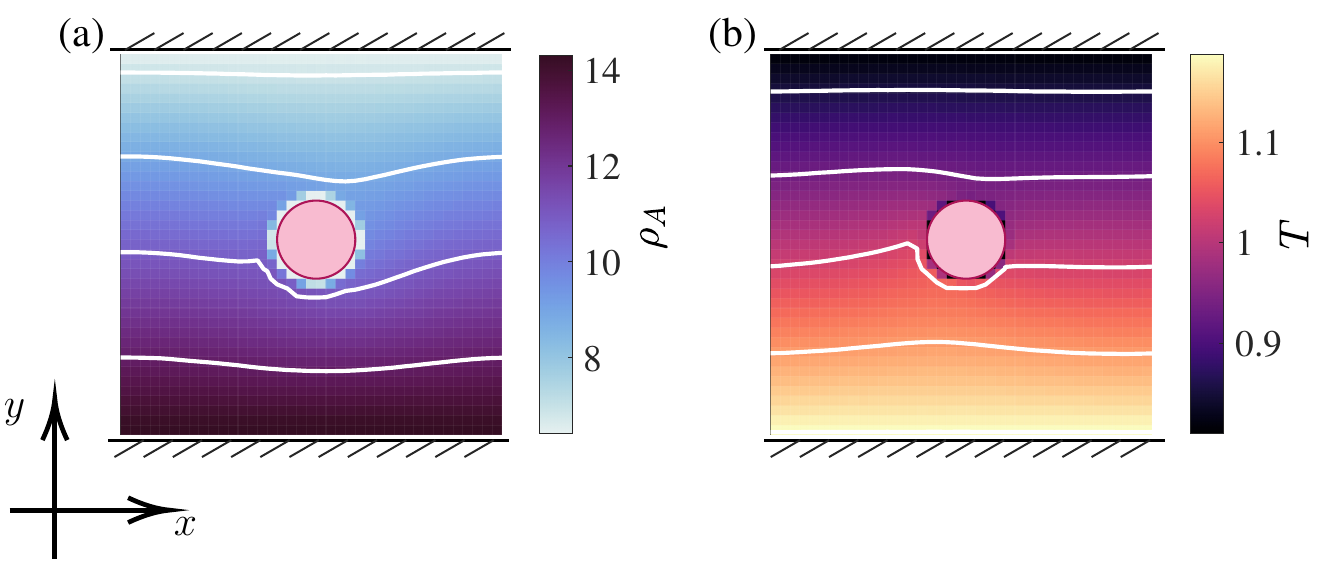}
    \caption{The established density/temperature field in simulations. These fields are linear in the far-field and distorted near the colloidal particle. The white lines indicate the isodensity/isotherms lines.}
    \label{Fig::Sketch}
\end{figure}

The simulation is performed in a square box with size $L=40$ (see Fig.~\ref{Fig::Sketch}). The periodic boundary condition is applied along the $x$-direction and the upper and bottom boundaries are no-slip walls. The colloidal particle is externally fixed at the center of the box: $\left( L/2, L/2 \right)$. The average densities of $A$ and $B$ fluid particles are set to $\overline{n}_A=\overline{n}_B=10$. To generate a density gradient (along the $y$-axis) of species $A$, we fix its density in two bins, i.e., $n_A\left( 0\leqslant y\leqslant 1  \right)=\overline{n}_A-\lambda \left( L-2 \right)/2$ and $n_A\left( L-1\leqslant y\leqslant L \right)=\overline{n}_A+\lambda \left( L-2 \right)/2$, where $\lambda$ is the density gradient. This is achieved by converting the species of each particle in these areas to species $A$ with the corresponding probability $p_A=\frac{n_A}{\overline{n}_A+\overline{n}_B}$ (every $40$ MD steps). Note that in our simulation, the total density is kept unchanged ($n=n_A+n_B=20$). After a relaxation, a density field with gradient of $\lambda$ for species $A$ is established along the $y$-axis (see Fig.~\ref{Fig::Sketch}.(a)).

\subsection{The setup of thermophoresis}
The simulations of thermophoresis involve only fluid particles $A$ and the colloidal particle. The simulation box, boundary conditions, and the positions of the colloidal particles are the same as those in the diffusiophoresis simulations. Here, the average density of fluid particles is $\overline{n} = 10$.

In order to generate a temperature gradient (also along the $y$-axis), we fix the temperature in two bins (similar to the case of the diffusiophoresis simulations): $T\left( 0\leqslant y\leqslant 1 \right)=\overline{T}-\gamma \left( L-2 \right)/2$ and $T\left( L-1\leqslant y\leqslant L \right)=\overline{T}+\gamma \left( L-2 \right)/2$, where $\gamma$ is the temperature gradient. In these two bins, after every collision step, we apply the Maxwell-Boltzmann scaling thermostat~\cite{MBS_Huang}, which is developed for the original SRD. The resulting steady-state temperature field is depicted in Fig.~\eqref{Fig::Sketch}.(b).

\section{Simulation results}\label{Results}
\subsection{Phoretic factor tensor}
The diffusiophoretic/thermophoretic factor tensor can be easily determined from Eqs.~\eqref{FT} and \eqref{FD} by computing the total force exerted by the fluid particles on the colloidal particle. It is worth noting that the CSRD fluid obeys the ideal gas equation of state~\cite{CSRD}, so the chemical potential in Eq.~\eqref{FD} is $\mu=k_BT\ln \left( n_A/n \right)$.  The calculated results are summarized in Table~\ref{Tab::Fcts}. For comparison, we also quantify $\bm{\alpha}^{\rm{D}}$ and $\bm{\alpha}^{\rm{T}}$ for a normal fluid ($\theta=0$) and an odd fluid with opposite chirality ($\theta=-5\pi/9$). Unlike the normal case ($\theta=0$), the odd diffusiophoretic and thermophoretic factors are nonvanishing in the odd fluid ($\theta=\pm 5\pi/9$). Furthermore, we observe that the sign of the odd phoretic factors reverses in a system with opposite chirality. These results explicitly confirm the existence of odd phoresis in our system.

\begin{table}[htbp]
    \caption{\label{Tab::Fcts}%
        Components of $\bm{\alpha}^{\rm{D}}$ and $\bm{\alpha}^{\rm{T}}$ measured from simulations. The thermodynamic gradients $\nabla n_A=\lambda\bm{e}_y$ and $\nabla T=\gamma\bm{e}_y$ are set as $\lambda=-0.2$ and $\gamma=-0.01$, respectively.
    }
    \begin{ruledtabular}
        \begin{tabular}{l||cc|cc}
            \multirow{2}{*}{Fluid type} &  {$\alpha^{\rm{D}}_e$}  & \multicolumn{1}{c|}{$\alpha^{\rm{D}}_o$} & $\alpha^{\rm{T}}_e$ & $\alpha^{\rm{T}}_o$ \\
                            & \multicolumn{2}{c|}{($\omega=2\pi/3$)} & \multicolumn{2}{c}{($\omega=5\pi/6$)} \\ \hline 
            $\theta=5\pi/9$ &  49.2(4) & 6.3(3)  & -43.4(2) & 0.9(1)  \\
            $\theta=0$      &  57.7(3) & 0        & -56.6(7) & 0           \\
            $\theta=-5\pi/9$&  49.2(4) & -6.1(3) & -43.2(2) & -0.9(1)  \\ 
        \end{tabular}
    \end{ruledtabular}
\end{table}

\subsection{Flow field induced by odd phoresis}
\begin{figure*}[htbp!]
    \centering
    \includegraphics[keepaspectratio, width=1.4\columnwidth]{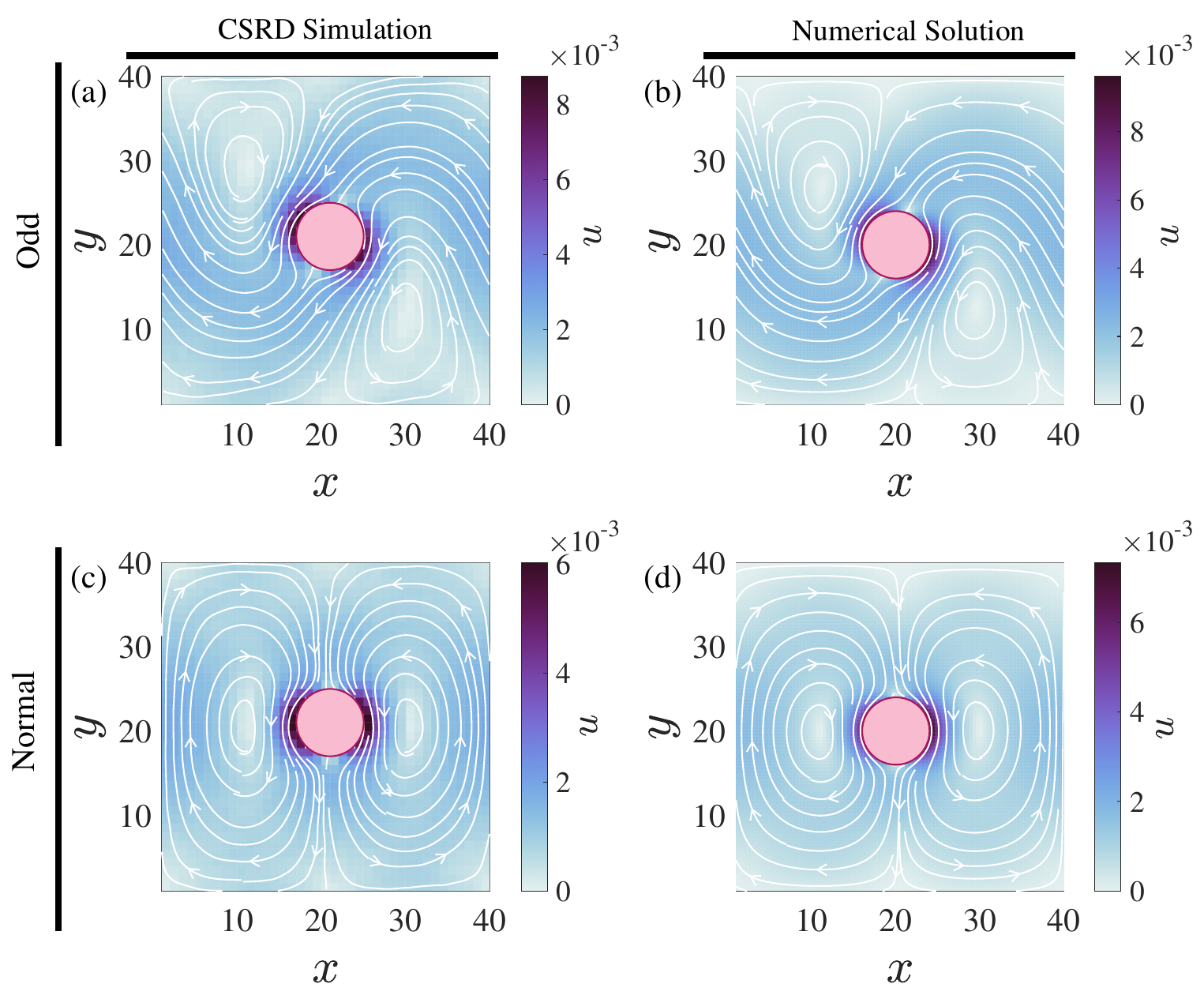}
    \caption{Flow fields induced by a fixed diffusiophoretic colloidal particle in odd fluid (a,b) and normal fluid (c,d). Here, (a,c) correspond to the particle-based simulations and (b,d) to the numerical solution of the Stokes equation. The background color represents the velocity modulus and the white lines are the flow streamlines. Viscosities used in numerical calculation are predicted by the kinetic theory~\cite{CSRD}: $\eta^{\text{kin}}\approx 2.60$, $\eta^{\text{col}}\approx 14.46$, $\eta_o^{\text{kin}}\approx-1.16$, $\eta_o^{\text{col}}\approx 7.80$ for the odd fluid ($\theta=5\pi/9$), and $\eta^{\text{kin}}\approx0.40$, $\eta^{\text{col}}\approx 23.75$, $\eta_o^{\text{kin}}=\eta_o^{\text{col}}=0$ for the normal fluid ($\theta=0$).}
    \label{Fig::PhFlows}
\end{figure*}

In this section, we investigate the flow field around an odd diffusiophoretic colloidal particle in the simulation. The flow fields generated by both odd and normal diffusiophoresis are depicted in Figs.~\ref{Fig::PhFlows}(a,c). Unlike the flow pattern of the normal fluid (Fig.~\ref{Fig::PhFlows}(c)), the odd diffusiophoretic osmotic flow with $\theta=5\pi/9$ (Fig.~\ref{Fig::PhFlows}(a)) exhibits broken mirror symmetry with respect to $x=L/2$. The flow pattern of an odd fluid with opposite chirality (i.e., the CSRD fluid with $\theta=-5\pi/9$) is the exact mirror image of the flow in Fig.~\ref{Fig::PhFlows}(a) across $x=L/2$.

{Because the flow is confined between no-slip boundaries, the Stokes paradox occuring in 2D unbounded fluids is bypassed by the truncation of the flow at the boundaries. Consequently, in the incompressible and low-Reynolds-number limit, the flow fields are well described by the Stokes equation for the CSRD fluid according to~\cite{CSRD}:}
\begin{equation}\label{Seq}
    \begin{aligned}
        &\bm{\nabla}\cdot\bm{u}=0,\\
        &\partial_\beta\sigma_{\alpha\beta}=0,
    \end{aligned}
\end{equation}
where
\begin{equation}\label{Stress}
    \begin{aligned}
        \sigma_{\a\b}=&\,-\tilde{p}\kr{\a\b}+\eta\left( \pd{\b}u_\a+\pd{\a}u_\b \right)\\
                      &+\eta_R\left( \pd{\b}u_\a-\pd{\a}u_\b \right)+2\eta_o\ep{\b\g}\pd{\g}u_\a.
    \end{aligned}
\end{equation}
Here, $\eta$ is the shear viscosity, $\eta_R$ is the rotation-rotation viscosity (see~\cite{Vitelli_Fluhydro} for its physical meaning), $\eta_o$ is the odd viscosity, $\tilde{p}=p-\hat{\eta}_o\ep{\a\b}\pd{\a}u_\b$ is the effective pressure, and $\hat{\eta}$ and $\hat{\eta}_o$ are combinations of viscosities. In the CSRD, each type of viscosity consists of a kinetic part (from the streaming step) and a collisional part (from the collision step):
\begin{equation}
    \begin{aligned}
        &\eta=\eta^{\text{kin}}+\frac{1}{2}\eta^{\text{col}},\qquad \eta_R=\frac{1}{2}\eta^{\text{col}},\\
        &\eta_o=\eta_o^{\text{kin}}+\frac{1}{2}\eta_o^{\text{col}},\\
        &\hat{\eta}=\eta^{\text{kin}}+\eta^{\text{col}},\qquad \hat{\eta}_o=\eta_o^{\text{kin}}+\eta_o^{\text{col}}.
    \end{aligned}
\end{equation}
{The values of these viscosities can be calculated from the analytical expressions derived in~\cite{CSRD}:
\begin{equation}
    \begin{aligned}
      \eta^{\text{kin}}=&\ \frac{nk_BT\h\n}{\n-1+e^{-\n}}\frac{1-\cos2\theta\cos2\omega}{1+\cos2\omega(\cos2\omega-2\cos2\theta)}\\
      &-\frac{1}{2}nk_BT\h,\\
      \eta^{\text{col}}=&\ \frac{m}{12\h}\left( \n-1+e^{-\n} \right)\left( 1-\cos\theta\cos\omega \right),\\
      \eta_o^{\text{kin}}=&\ -\frac{nk_BT\h\n}{\n-1+e^{-\n}}\frac{\sin2\theta\cos2\omega}{1+\cos2\omega(\cos2\omega-2\cos2\theta)},\\
      \eta_o^{\text{col}}=&\ \frac{m}{12\h}\left( \n-1+e^{-\n} \right)\sin\theta\cos\omega,
    \end{aligned}
\end{equation}
where $n$ is the particle number density and $\n$ is the mean particle number in a collision cell of CSRD.}

The boundary conditions for the colloidal diffusiophoresis are as follows: 1). no-slip upper and lower boundary walls; 2). periodic boundary condition along $x$-axis; 3). nonvanishing diffusiophoretic force acting on the colloidal particle. These conditions are explicitly written as:
\begin{equation}\label{Bnd}
    \begin{aligned}
        &\left.\bm{u}\right|_{y=0,L}=\bm{0},\\
        &\bm{u}\left( x,y \right)=\bm{u}\left( x+L,y \right),\\
        &\oint_S \bm{\sigma}\cdot d\bm{n}=\bm{F}^{\rm{D}},
    \end{aligned}
\end{equation}
where $S$ is the surface of the colloidal particle with normal vector $\bm{n}$ (the colloidal particle is a disk with radius $\sigma=4$). We solve this boundary problem (Eqs.~\eqref{Seq} and \eqref{Bnd}) by finite element method (FEM) for both normal fluid and odd fluid (see Figs.~\ref{Fig::PhFlows}(b,d)). {In FEM, the diffusiophoretic force $\bm{F}^{\rm D}$ is taken directly from the measurements in the CSRD simulations of diffusiophoresis, and all viscosities in~\eqref{Seq} are set to the values according to the CSRD simulations.} These numerical solutions show an excellent agreement with the CSRD simulation results.

{
\section{Discussion}
\subsection{Finite-size effects}
Herein, we provide an analysis of finite-size effects in diffusiophoresis simulations. Generally speaking, in the current confined system, the back-flow effect generated by the walls can affect the phoretic force acting on the colloid, which should deviate from that in the unbounded fluid where the system size $L$ goes to infinity. More concretely, the back-flow reflected from the bottom boundary is longitudinal and points in the same direction as the longitudinal phoretic force $F^{\text{D}}_y$, thereby enhancing $F^{\text{D}}_y$. Meanwhile, this back-flow is sheared by the colloidal particle, leading to a positive shear rate $\partial_xu_y$ on the right side of the colloidal particle and a negative one on the left. According to the CSRD stress constitutive relation Eq.~\eqref{Stress}, the shear rate $\partial_xu_y$ contributes to the normal stress $\sigma_{xx}$ through $\sigma_{xx}=-p+\hat{\eta}_o\partial_xu_y$. Therefore, the difference in the shear rate between the left and right sides of the colloidal particle results in a transverse lift force acting on the particle (see also~\cite{Lou2022} for a similar mechanism underlying the lift force). When $\hat{\eta}_o>0$, this lift force points to the right, in the same direction as the transverse phoretic force measured in the simulation, and thus enhances the transverse phoretic force. When $\hat{\eta}_o<0$, both the lift force and the transverse phoretic force reverse direction, so the lift force still remains aligned with the transverse phoretic force. Consequently, we conclude that the phoretic force is enhanced in a finite-size system. A similar analysis is also applicable for the thermophoresis.

Thus, we expect the diffusiophoretic factors $\alpha_e^{\text{D}}$ and $\alpha_o^{\text{D}}$ to be smaller in free space compared to those in a confined space. To quantitatively estimate the free-space values of these factors, denoted by $\alpha_e^{\text{D},\infty}$ and $\alpha_o^{\text{D},\infty}$, we performed simulations to measure $\alpha_e^{\text{D}}(L)$ and $\alpha_o^{\text{D}}(L)$ in systems with different sizes under the same concentration gradients: $L=30,\,40,\,50,\,60,\,80$, and applied the following linear fitting relations
\begin{equation}
    \alpha_e^{\text{D}}(L)=\alpha_e^{\text{D},\infty} + k_e\frac{\sigma}{L},\qquad\alpha_o^{\text{D}}(L)=\alpha_o^{\text{D},\infty} + k_o\frac{\sigma}{L}
\end{equation} 
to extract $\alpha_e^{\text{D},\infty}$ and $\alpha_o^{\text{D},\infty}$. As shown in Figs.~\ref{Fig::FS}(a) and (b), these linear expressions fit well with the original data. The extrapolated $\alpha_e^{\text{D},\infty}$ and $\alpha_o^{\text{D},\infty}$ are:
\begin{equation}
\alpha_e^{\text{D},\infty}=12.2\pm 0.9,\qquad\alpha_o^{\text{D},\infty}=1.9\pm 0.6.
\end{equation}
We note that the odd diffusiophoretic factor remains nonvanishing in the limit $L\to\infty$ considering the margin of error, confirming the existence of the odd diffusiophoresis. Although finite-size scaling for odd thermophoresis was not performed due to the higher computational cost, we expect it will also remain nonvanishing in the limit $L\to\infty$. This is because the intrinsic parity-breaking property of the CSRD fluid can lead to a parity-breaking fluid particle distribution around the colloidal particle, consequently generating a transverse phoretic force on the colloid.

\begin{figure}[th]
    \centering
    \includegraphics[keepaspectratio, width=\columnwidth]{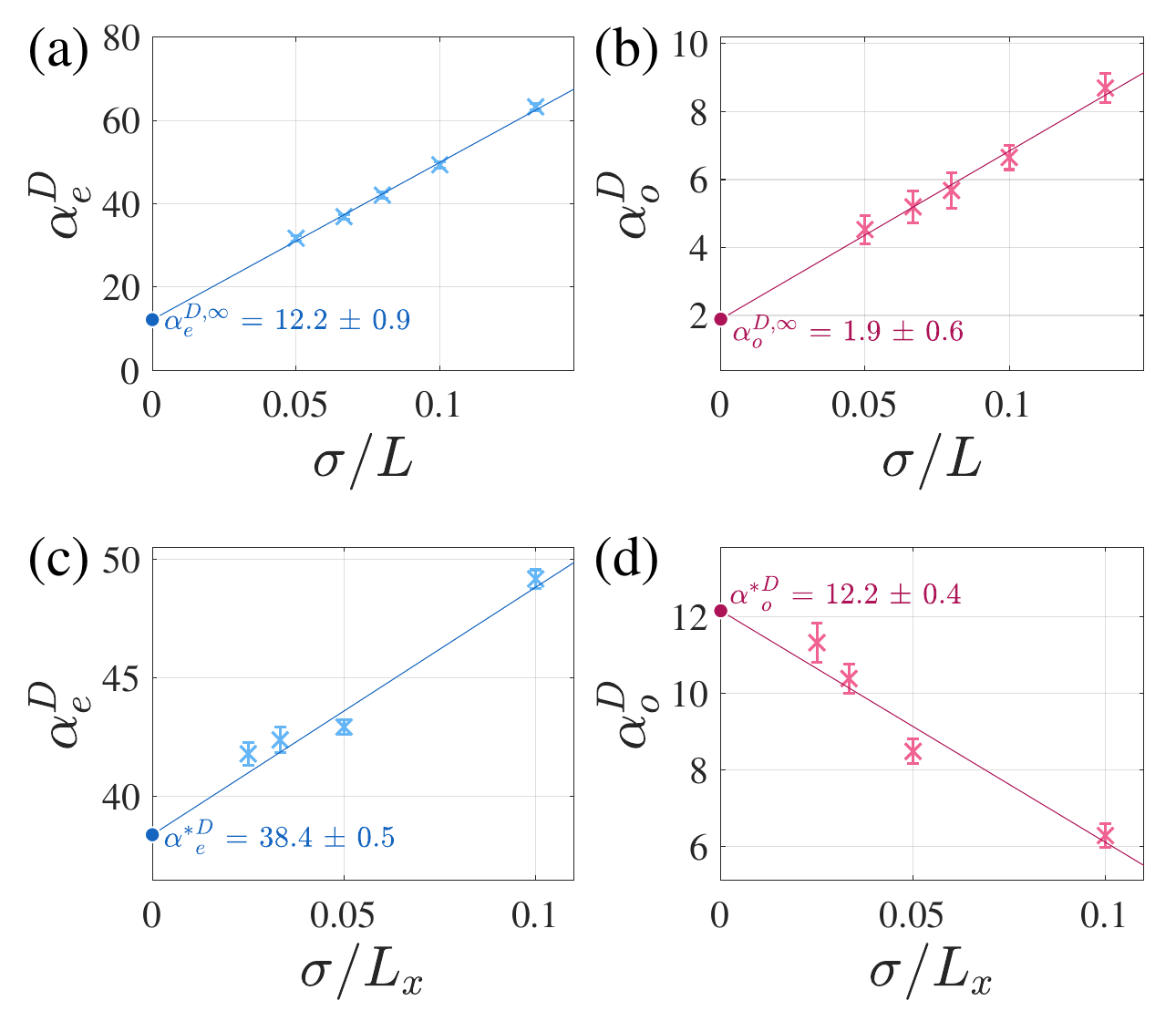}
    {\caption{(a)-(b) Finite-size scaling of $\alpha_e^{\text{D}}$ and $\alpha_o^{\text{D}}$. The linear fits are given by the solid lines, with fitting parameters $\alpha_e^{\text{D},\infty}=12.2\pm 0.9$, $k_e=(3.7\pm0.1)\times 10^2$ and $\alpha_o^{\text{D},\infty}=1.9\pm 0.6$, $k_o=50\pm 6$, respectively. The density gradient here is set to $\lambda=-0.1$, and other parameters remain unchanged: $\epsilon=2.5$, $\sigma=4$, $\h=50\Delta t_{MD}=0.1$, $\theta=5\pi/9$, $\omega=2\pi/3$. (c)-(d) Measured values of $\alpha_e^{\text{D}}$ and $\alpha_o^{\text{D}}$ in systems with different $L_x$.The linear fits are also shown by the solid lines, with fitting parameters ${\alpha_e^*}^D=38.4\pm 0.4$, $s_e=(1.04\pm0.07)\times 10^2$ and ${\alpha_o^*}^D=12.0\pm 0.4$, $s_o=-60\pm 6$, respectively. The longitudinal size is fixed at $L_y=40$, and other parameters remain the same: $\lambda=-0.2$, $\epsilon=2.5$, $\sigma=4$, $\h=50\Delta t_{MD}=0.1$, $\theta=5\pi/9$, $\omega=2\pi/3$.}\label{Fig::FS}}
\end{figure}

\subsection{Influence of the net transverse flow}
Finally, we discuss a subtle approximation used in the determination of odd phoretic factors. Because of the periodic boundary condition in the $x$ direction, a net transverse flow is induced by the reaction of the odd phoretic force in the odd fluid. This net transverse flow in turn applies frictional forces on both the no-slip walls and the colloidal particle. In the above analysis, the frictional force on the colloidal particle has been implicitly neglected. As a result, we used the total force (including the phoretic force and partial friction) exerted by the fluid on the colloidal particle to approximate the phoretic force. In fact, the transverse frictional force on the particle acts in the opposite direction to the odd phoretic force, thus reducing the total transverse force on the colloidal particle exerted by the fluid. Consequently, the true odd phoretic force should be even stronger than the present measured value, which does not qualitatively change our conclusion.

To examine the influence of this net transverse flow, we can increase the length in the periodic direction of the system (i.e., the transverse size, denoted by $L_x$) to enhance the wall friction and weaken the transverse flow, while keeping the longitudinal size $L_y$ fixed. Under the weaker net transverse flow, the total force acting on the colloid becomes closer to the real phoretic force and the measured phoretic factors are thus closer to their real values. 

As an example, we further determine the real values of the diffusiophoretic factors through a linear extrapolation of the diffusiophoretic factors measured in systems with different transverse sizes $L_x$ (but $L_y=L$ and the concentration gradient remains fixed). Specifically, we denote the real values of diffusiophoretic factors by ${\alpha_e^*}^D$ and ${\alpha_o^*}^D$, and the measured values by $\alpha_e^{\text{D}}(L_x)$ and $\alpha_o^{\text{D}}(L_x)$. We assume the following linear relations for large $L_x$:
\begin{equation}
  \alpha_e^{\text{D}}(L_x)={\alpha_e^*}^D + s_e\frac{\sigma}{L_x},\qquad\alpha_o^{\text{D}}(L_x)={\alpha_o^*}^D + s_o\frac{\sigma}{L_x}.
\end{equation}

By performing the CSRD simulations of $L_x\times L_y$ systems with fixed $L_y=L=40$ and different $L_x=40,\,80,\,120,\,160$, we obtain the corresponding $\alpha_e^{\text{D}}(L_x)$ and $\alpha_o^{\text{D}}(L_x)$. The simulation results and corresponding linear fits are shown in Figs.~\ref{Fig::FS}(c) and (d). We find that the measured odd diffusiophoretic factor $\alpha_o^{\text{D}}(L_x)$ increases, whereas the normal diffusiophoretic factor $\alpha_e^{\text{D}}(L_x)$ decreases, with increasing $L_x$. Both of these effects are induced by the net transverse flow. First, $\alpha_o^{\text{D}}(L_x)$ increases because the transverse friction (against the transverse diffusiophoretic force) exerted on the colloid by the net flow diminishes. Secondly, because the transverse flow around the colloidal particle is sheared, the odd viscosity can generate a lift force whose direction is aligned with the longitudinal diffusiophoretic force, as discussed above. This implies the measured normal diffusiophoretic factor $\alpha_e^{\text{D}}(L_x)$ is larger than the real value. Therefore, $\alpha_e^{\text{D}}(L_x)$ decreases as $L_x$ increases.

}

\section{Conclusion}\label{Conclusion}
With the help of CSRD, a mesoscopic model for odd fluids, we successfully realize the 2D odd diffusiophoresis and thermophoresis of a colloidal particle, which have not been reported in either experiments or simulations before. The corresponding phoretic factor tensors are determined. We further measure the flow field induced by a fixed odd diffusiophoretic colloidal particle in the CSRD simulations, which is consistent with the numerical solutions of the Navier-Stokes equation. These results demonstrate the existence of the odd cross transport phenomena in odd colloids and the power of the CSRD model for simulating odd complex fluids. Therefore, by using CSRD, we can further study other odd cross transports, a valuable but still relatively unexplored field~\cite{Mandadapu2024}. We also hope this work will motivate additional studies on odd phoresis, particularly its potential applications and exotic behavior in three-dimensional systems.

\section*{Acknowledgment}
This work was supported by the National Key R\&D Program of China
(2022YFF0503504) and the National Natural Science Foundation of China (12274448, T2325027, 12504241).

\bibliographystyle{apsrev}
\bibliography{RefList}

\end{document}